\documentclass[11pt]{article}
\usepackage{moriond,epsfig}
\def\Journal#1#2#3#4{{#1} {\bf #2}, #3 (#4)}

\def\PLB{{\em Phys. Lett.}  B}
\def\PRL{\em Phys. Rev. Lett.}

\def\APJ{\em Astrophys. J}

\def\be{\begin{equation}}
\def\ee{\end{equation}}
\def\bea{\begin{eqnarray}}
\def\eea{\end{eqnarray}}

\def\epm{$e^\pm $}

\begin{document}
\vspace*{4cm}
\title{THE COSMIC-RAY POSITRON AND ELECTRON EXCESS: AN EXPERIMENTALIST'S POINT OF VIEW}

\author{ M.~SCHUBNELL }

\address{Department of Physics, University of Michigan, 450 Church Street,\\
Ann Arbor, MI, USA}

\maketitle\abstracts{
The recent report by the PAMELA team of the observed rise in the cosmic-ray positron fraction above a few GeV and the report of an excess of cosmic-ray electrons around a few hundred GeV by the ATIC collaboration has resulted in a flurry of publications interpreting these observations either as a possible signature from the decay of dark matter or as a contribution from isolated astrophysical sources. While those interpretations are scientifically exciting, the possibility that measurements are contaminated by misidentified cosmic ray protons can not be ignored.}

\section{Introduction}
Positrons and electrons constitute only a small fraction of the cosmic-ray intensity measured near earth but the observation of these particles has long been recognized as a powerful tool for the investigation of cosmic-ray production and acceleration as well as transport and interaction in the interstellar medium. Direct observation of the all-electron (electrons {\it and} positrons, \epm) component is feasible up to energies of about a few TeV by balloon or satellite experiments while at higher energies the low particle fluxes require indirect techniques with effective detection areas much larger than the instrument itself.

From the measurement of the relative abundances of positrons and electrons it is evident that the all-electron component consists of mainly electrons. These are generally thought to originate from the same sources as the hadronic component of the cosmic radiation but unlike the much more massive hadrons, electrons experience severe energy loss during propagation, dominantly through synchrotron losses and inverse Compton scattering on interstellar photons. Above GeV energies $\le$ 10\% of the electrons in the cosmic radiation and an nearly equal amount of positrons are produced through the decay of secondary particles, mostly pions, generated in hadronic interactions of energetic protons with the interstellar medium.
The relative fraction of positrons, {$N_{e^+}/(N_{e^-} + N_{e^+})$},
is expected to fall slowly
with energy because of the declining path length of the primary
nuclei at high rigidities. The pure secondary nature of the positron component makes it a good probe of cosmic-ray propagation and the precise understanding of this constituent of the all-electron intensity allows to isolate the characteristics of the primary electron component. 

The absolute intensities of cosmic ray protons and electrons has been measured extensively and can, in the energy region between about 1 GeV and 100 GeV, be represented by a power-law of the form $I(E)=b \times E^{-\alpha}$, with the spectral index {$\alpha_p \approx 2.7$} for protons and  {$\alpha_e \approx 3.3$} for electrons.

\section{Cosmic-Ray Positron and Electron Measurements}

\begin{figure}
\vskip 0.5cm
\begin{minipage}[b]{0.5\linewidth} 
\centering
\psfig{figure=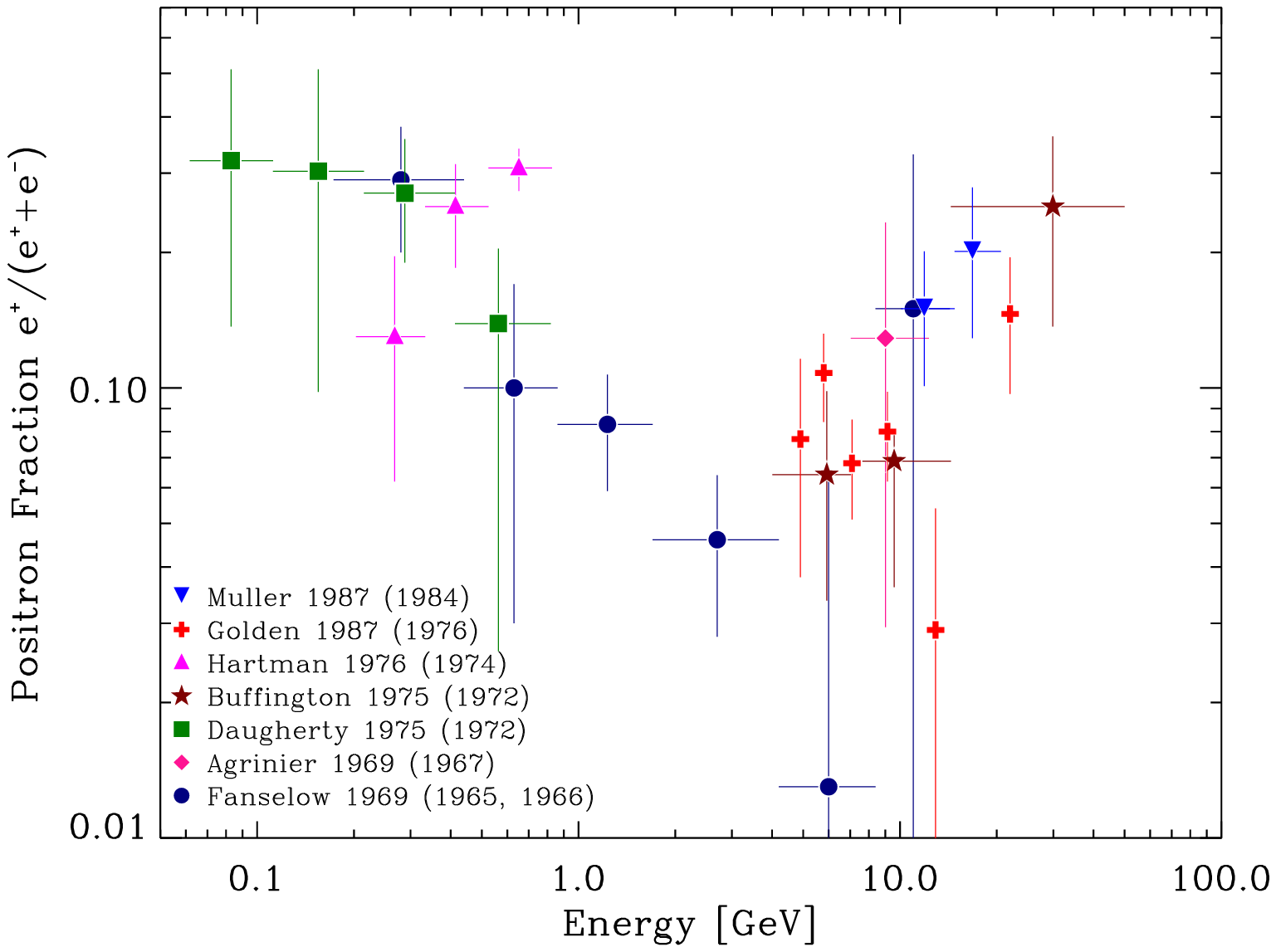,width=1.0\textwidth}
\caption{\protect \raggedright Measurements of the positron fraction prior to the 1990s. The year(s) the data was taken is indicated in parenthesis. For references see Barwick {\it et al} {\protect \cite{barwick94}}. }
\label{fig:positron_fraction_old}
\end{minipage}
\hspace{0.2cm}
\begin{minipage}[b]{0.5\linewidth}
\centering
\psfig{figure=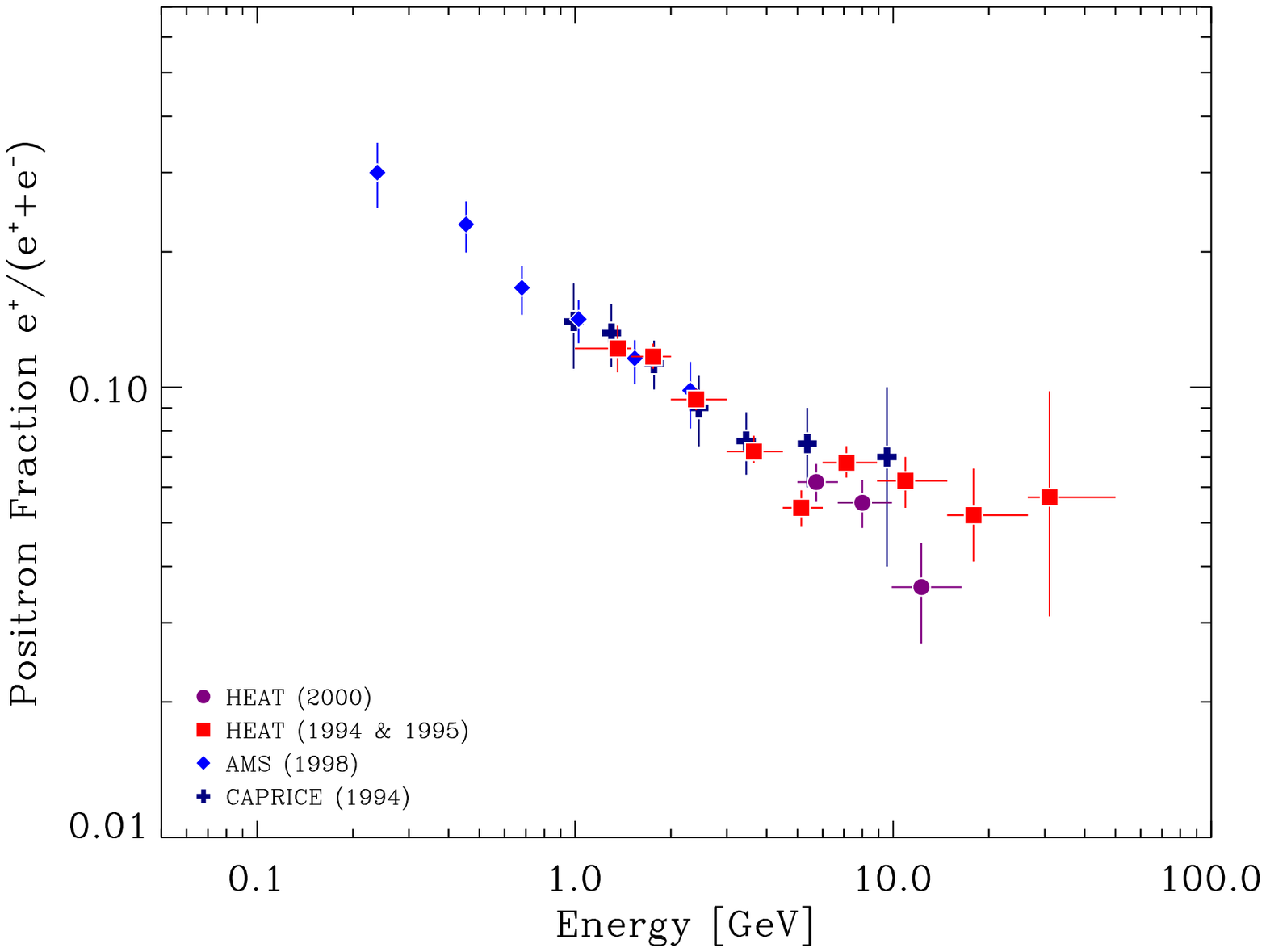,width=1.0\linewidth}
\caption{\protect \raggedright Recent measurements of the positron fraction  do not confirm the rise seen in earlier data {\protect \cite{barwick94,beatty,boezio,alcaraz}}. The HEAT group reports a possible structure at $\approx$7~GeV \protect \cite{barwick94}.}
\label{fig:positron_fraction_new}
\end{minipage}
\end{figure}

\begin{figure}
\vskip 0.5cm
\centering
\psfig{figure=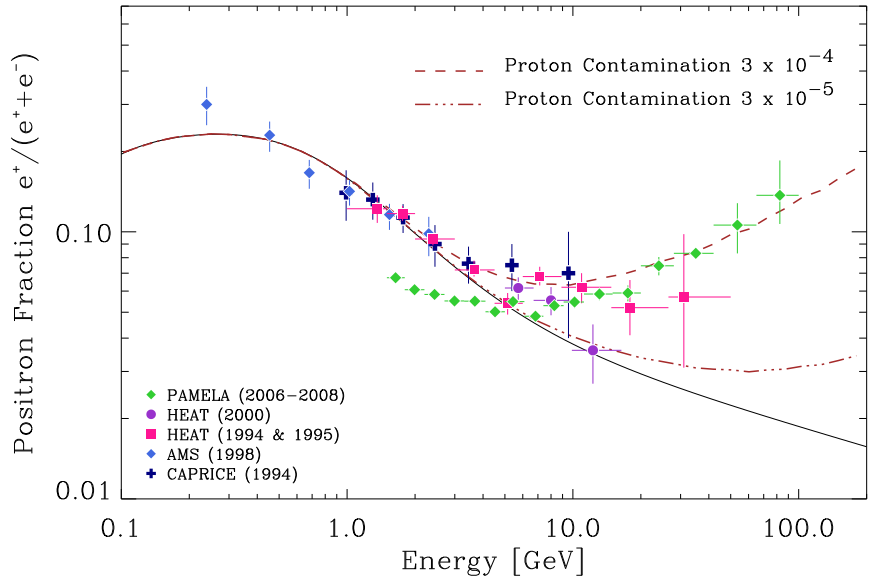,height=3.0in}
\caption{\protect \raggedright Recent measurements of the positron fraction overlaid with a pure secondary production prediction for cosmic-ray positrons {\protect \cite{moskalenko}} and the same prediction including residual proton contamination. Below $\approx$5 GeV solar modulation affects the particle intensities observed near Earth and may explain the discrepancy between the PAMELA data and older measurements, obtained at distinctively different solar epochs {\protect \cite{clem}}. In the region between 5 and 50 GeV measurements by PAMELA are consistent with previous data from the HEAT experiment, while the dramatic rise at high energies is reminiscent of an earlier era when particle identification was insufficient.}
\label{fig:positron_fraction}
\end{figure}

The majority of early cosmic-ray positron measurements (prior to the 1990s), mostly utilizing small permanent magnet spectrometers and detector systems with limited particle identification capability, observed an unexpected rise in the positron fraction above $\approx$5 GeV (Fig. \ref{fig:positron_fraction_old}). Subsequent measurements by instruments with more powerful particle identification showed the almost exclusively secondary nature of positrons with no evidence for a steeply rising positron fraction (Fig. \ref{fig:positron_fraction_new}).
At the same time a possible structure near 7 GeV was reported by the {HEAT-\epm} collaboration \cite{barwick94}. These results were confirmed by the CAPRICE instrument \cite{boezio} and the AMS-1 instrument \cite{alcaraz}, albeit with lower resolution and energy reach. 
The unexpected feature in the positron fraction was discussed by many authors as a possible signature of dark matter. Very recently, the PAMELA satellite experiment extended the cosmic-ray positron and electron observations to higher energies \cite{adriani}, confirming the measurements by HEAT but also claiming a dramatic rise in the positron fraction starting at 10 GeV and extending up to 100 GeV, reminiscent of the rise observed at lower energies by early observations (Fig. \ref{fig:positron_fraction}).  At yet higher energies, around a few 100 GeV, the ATIC instrument reports a significant excess in the all-electron intensity \cite{chang}. The PAMELA and ATIC measurements have motivated a great number of interpretations, from possible signatures of dark matter to astrophysical sources \footnote{An up-to-date selection can be found at the astrophysics archive: http://arxiv.org/archive/astro-ph.}. While such explanations are very attactive and exciting one has to use caution when interpreting cosmic-ray positron and electron measurements.

The intensity of cosmic-ray protons at 10 GeV exceeds that of positrons by a
factor of about $5\times10^4$. Therefore a proton rejection of about $10^6$ is required if one wants to obtain  a positron sample with less than $\approx$5\% proton contamination. Thus, from an experimental point of view, the single biggest challenge in measuring cosmic-ray positrons is the discrimination against the vast proton background. Furthermore, because the proton spectrum is much harder than the electron and positron spectrum the proton rejection has to improve with energy. In addition, any small amount of spillover from tails in lower energy bins can become problematic.

Keeping this in mind it is easy to see why proper particle identification is crucial for the measurements and becomes more important at higher energies. Excellent particle identification requires multiple detector systems and techniques, ideally redundant and complementary, allowing for in-situ determination of hadronic rejection efficiency. Such an instrument will typically employ a powerful magnet spectrometer which measures the particle's charge sign and rigidity. The higher the magnetic field strength and the finer the granularity of the hodoscope's tracking layers the higher the rigidities that can be reached. The particle's  charge and direction is typically measured by `time of flight' scintillator layers. Hadron/electron separation is achieved with transition radiation detectors (TRDs) which measure the ratio of a particle's mass and energy and therefore are ideal for this purpose.  
Finally, an electromagnetic calorimeter below the magnet spectrometer results in particle identification and energy determination through shower shape and electromagnetic energy deposition. Because of weight restrictions, for balloon borne and satellite instruments the calorimeter depth is typically limited to 10-15 radiation lengths.

The PAMELA instrument suffers from the lack of a second, powerful hadron/electron discrimination detector element (such as a TRD) and therefore relies on the electromagnetic calorimeter for hadron rejection.  This impacts positron measurements because with increasing energy not only does the proton background increase but also the discrimination of electromagnetic showers inside the calorimeter becomes more difficult. The probability that hadronic particles mimic electromagnetic showers through early $\pi^0$ production is problematic. Very small amounts of spill-over from lower energy bins adds to the list of potential problems. The neutron detector on PAMELA can come to the rescue at high energies  but is inefficient below about 100 GeV.  However, in the case of early $\pi^0$ production with no further hadronic interaction, the neutron detector will be inefficient. 

The published all-electron measurements by the ATIC group are difficult to reconcile with previous measurements. The claimed excess has given rise to numerous speculative publications interpreting the data as evidence for primary electrons from dark matter decay. However, from an experimentalist's perspective, the reported data are suspicious as the authors do not properly take into account the uncertainties associated with potential hadronic background particle interactions inside the graphite target on top of the detector. Indeed, this experiment was designed and optimized to detect hadronic particles and while the low Z target is good for detecting nuclei it increases the probability of hadronic contamination. Additionally, uncalibrated leakage of the electromagnetic shower out the back of the calorimeter can lead to pile-up at lower energies.

\section {Conclusions and Outlook}

Cosmic-ray electron and positron measurements are uniquely capable of illuminating the century old question of the origin of cosmic rays. Electromagnetic process are well understood (at least at the energies in question) and experiments, while challenging, can precisely measure the two components to energies up to at least 100 GeV. The HEAT-\epm \/ instrument has shown that the combination of powerful detector systems can successfully suppress the hadronic background to the required level.  The data reported by PAMELA are, if correct, exciting and confirm the HEAT results of an excess in the positron fraction above a few GeV. While cosmic-ray experiments are and will continue to be important in searching for possible dark matter signatures in cosmic rays, it has to be emphasized that the observation of such an excess is not necessarily linked to dark matter and may well reflect the contribution of primary astrophysical sources to the positron flux. Any potential dark matter `signature' observed in cosmic rays will ultimately rely on confirmation through accelerator data.  

Caution should be exercised when interpreting cosmic-ray positron and electron data above a few GeV because of possible proton contamination of the measurements. The ATIC results in particular are suspicious and will be scrutinized by the forthcoming results from FERMI.
Balloon borne instruments have been pathfinding in measuring cosmic-ray particles and have produced the majority of currently available data. While space borne instruments have the advantage of much longer exposure, they have to overcome the technical, financial, and often political hurdles to bring such an instrument into space. The AMS team for instance has been struggling for a decade to get a shuttle flight for deployment on the International Space Station. At costs in excess of \$1.5 billion one must ask if a similar size balloon borne instrument flown on several long duration balloon flights would not have been scientifically and financially more prudent. Typical long duration flights from McMurdo, Antarctica achieve now in excess of 40 days at float altitude. A single such flight would give an experiment like HEAT\epm \/ roughly 3 years worth of PAMELA data during a single flight. 
At energies above a few TeV, the particle rates become so small that the direct detection of cosmic-ray electrons and positrons becomes difficult at best. Indirect observations with ground based instruments such as HESS \cite{aharonian} will extend the energy reach to $\approx$5-10 TeV but with large systematic errors. Above 10 TeV the CREST instrument has the potential to measure the electron spectrum and to detect isolated sources of cosmic rays \cite{schubnell}.



\section*{References}
\begin{thebibliography}{99}

\bibitem{adriani}O. Adriani {\it et al}, \Journal{Nature}{458}{607}{2009}.

\bibitem{aharonian}F. Aharonian {\it et al}, \Journal{\PRL}{101}{261104}{2008}.

\bibitem{alcaraz}J. Alcaraz {\it et al}, \Journal{\PLB}{484}{10}{2000}.

\bibitem{barwick94}S.W. Barwick {\it et al}, \Journal{\APJ}{482L}{191}{1997}.

\bibitem{beatty}J.J. Beatty {\it et al}, \Journal{\PRL}{93}{241102}{2004}.

\bibitem{boezio}G. Barbiellini{\it et al}, \Journal{\em A\&A}{309}{L15}{1996}.

\bibitem{chang}J. Chang {\it et al}, \Journal{Nature}{456}{362}{2008}.

\bibitem{clem}J. Clem {\it et al}, \Journal{\APJ}{464}{507}{1996}.

\bibitem{duvernois}M.A. DuVernois {\it et al}, \Journal{\APJ}{559}{296}{2001}.

\bibitem{moskalenko}I. Moskalenko \& A. Strong, \Journal{\APJ}{493}{694}{1998}.

\bibitem{schubnell}M. Schubnell {\it et al}, \Journal{\em Proc. 30th ICRC, Merida, Mexico} {2}{305}{2007}.

\end{thebibliography}
\end{document}